# ZD-AOMDV: A New Routing Algorithm for Mobile Ad-Hoc Networks

Nastooh Taheri Javan
Computer Engineering Department
Amirkabir University of Technology
Tehran, Iran
nastooh@aut.ac.ir

Reza Kiaeifar
Azad University
Branch of Andimeshk
Andimeshk, Iran
kiaeefar@gmail.com

Bahram Hakhamaneshi
Computer Engineering Department
California State University
Sacramento, USA
hakhamab@ecs.csus.edu

Mehdi Dehghan
Computer Engineering Department
Amirkabir University of Technology
Tehran, Iran
dehghan@aut.ac.ir

***Abstract*— A common characteristic of all popular multi-path routing algorithms in Mobile Ad-hoc networks, such as AOMDV, is that the end to end delay is reduced by utilization of parallel paths. The competition between the neighboring nodes for obtaining a common channel in those parallel paths is the reason for end to end delay increment. In fact, due to medium access mechanism in wireless networks, such as CSMA/CA, data transmissions even through two Node-Disjoint paths are not completely independent and each path will affect the other one. In this paper we have modified the AODV protocol which results in selection of zone-disjoint paths, to the extent feasible, and as a result we achieve less end to end delay. The efficiency of the proposed protocol has been evaluated on different scenarios and there has been a noticeable improvement in the packet delivery ratio and also in the reduction of end-to-end delay comparing to AOMDV.**

## I. INTRODUCTION

In ad-hoc networks, lack of any infrastructure requires the nodes to perform routing and transferring of data all by themselves [1]. Mobility of nodes and ambiguity of topology adds to the complexity of routing in such networks [2].

There are many common routing algorithms used in Ad-hoc networks but AODV[1] [3, 4] and DSR[2] [5], which both are on-demand algorithms, are the most popular ones. In on-demand algorithms, the Route Discovery procedure is carried out only when there is a packet to be transferred and there exists no valid path. Multi-path on-demand routing algorithms discover several paths instead of one, once the routing is performed [6]. This eliminates the need for further routing when there is a broken link in the path, reducing the average number of Route Discovery for each node and achieving higher fault tolerance for the Mobile Ad-hoc networks.

In some multi-path algorithms once different paths are discovered, they are all stored but only one of them is used for transferring the data. The other stored paths will become useful once the current one is broken. There are also other multi-path algorithms that transfer data over all discovered paths concurrently which reduces end to end delay and increases end to end bandwidth.

One important factor in latter kind of algorithms is to choose the best paths for load balancing data transfer. In this case, set of paths that have no common node among themselves will achieve highest fault tolerance, since a broken node will only affect one path. This type of paths are knows as Node Disjoint.

In wireless networks CSMA/CA [7] protocol is used for acquiring channel access and to prevent Hidden and Exposed Terminal problems. In this protocol, RTS[3] and CTS[4] packets sent forth and back between nodes force some nodes to wait until they can take part in next competition round for acquiring the channel. This will increase overall end to end delay in node-disjoint paths.

As an example, figure1 shows a hypothetical wireless network which is consisted of 10 nodes. In this figure the wireless transmission range of each node is shown. Also, nodes that are within transmission range of each other are connected through links.

In this figure, there are two Node Disjoint paths between nodes S and D which are S-I1-I2-I3-I4-D and S-I5-I6-I7-I8-D. Transmission of data over these paths is not independent of each other. The end-to-end delay of each path is actually dependant on the traffic volume on the other path and this is due to the RTS and CTS messages which are used by nodes in the network to prevent collision and Hidden and Exposed Terminal problems. As a result a node in either of those two paths might have to delay its data transmission due to receiving a CTS message from a node on the other path.

---

[1] Ad hoc On-demand Distance Vector
[2] Dynamic Source Routing
[3] Request To Send
[4] Clear To Send

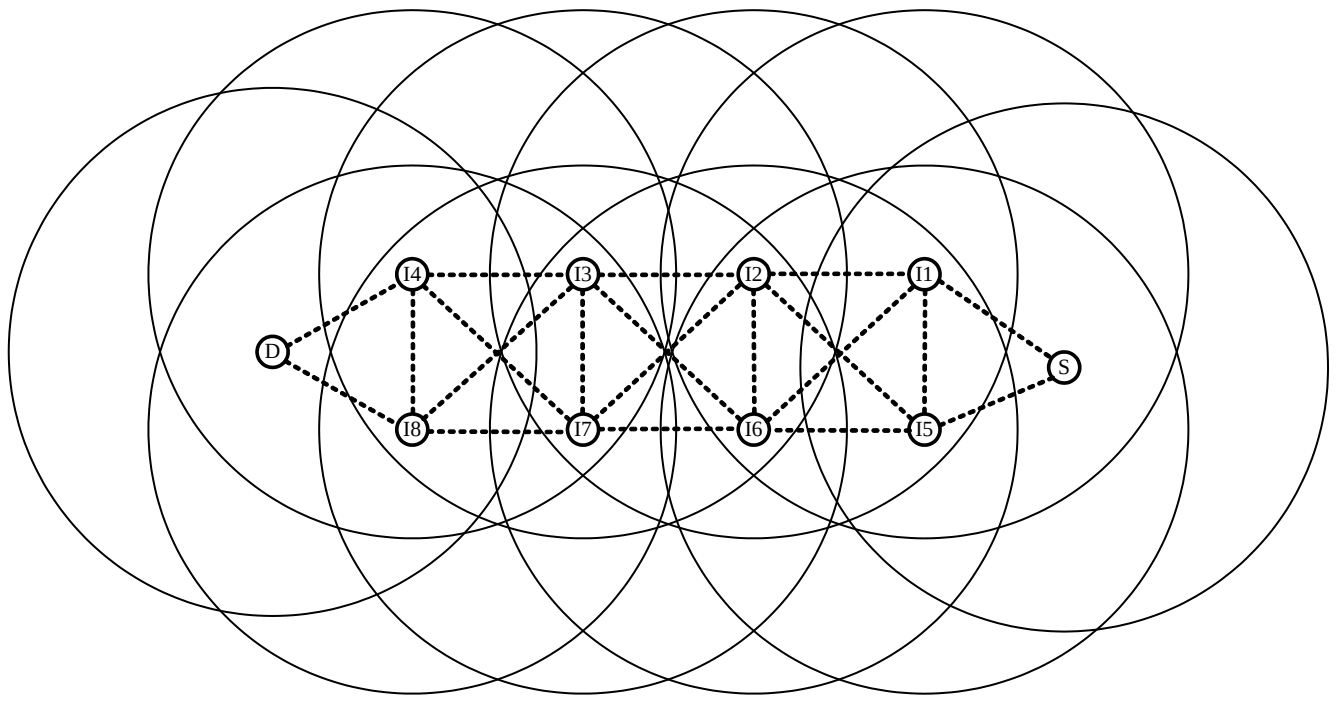

Figure 1. Node Disjoint Paths.

In this paper we propose an on-demand multi-path routing algorithm based on AODV, which utilizes common omni-directional antennas, unlike other solutions which require unidirectional antennas for each node in the network [8, 9].

The rest of this paper is organized as follows. The following section deals with the related works. Section III describes the proposed protocol mechanism in detail. Performance evaluation by simulation is presented in section IV and concluding remarks are made in section V.

## II. Related Works

Most of multi-path routing algorithms such as AOMDV[5] [10] have their basis in a single-path algorithm due to their efficiency. One drawback with multi-path algorithms is that they send more routing request packets, comparing to single-path algorithms and this makes them more complicated.

AOMDV is a simple, yet high efficient multi-path algorithm which tries to find Link-Disjoint paths. In this algorithm the source broadcasts routing request packet to all its neighbors. Once neighbors of the source node receive the RREQ packet, they add their own address to the packet; announcing themselves as the founder of a path, and broadcast it.

Every intermediate node by receiving a request packet will check to see if it has already received a same request and if not, then it reconstructs and broadcasts the request packet. If the same request was processed previously the following conditions will be investigated and if met, the packet would be accepted and will be inserted in the route table; otherwise the packet will be ignored. These conditions are:

1. Whether the request packet is received from a new neighbor node[6].
2. Whether the request packet is received from a new path founder node.
3. Whether the number of hops in the request packet is less than the existing ones.

The destination node will send as many responses as number of request packets it has received and hereby informs the source node of the paths.

To ensure zone-disjoint paths, [8] has utilized directional antennas. In this method each node stores signal power and signal angle of their neighbors in a table. While node-disjoint paths are being discovered, the information in that table will be used by each node to select zone-disjoint paths. One of the drawbacks of this method as mentioned in [9] is the need for directional antennas while in most ad-hoc network Omni-directional antennas are used.

## III. Proposed Algorithm

Our proposed algorithm, called ZD-AOMDV[7], is an on-demand multi-path routing protocol based on AODV. In the proposed algorithm the concept of "Active Neighbor" is introduced. Active neighbors are the neighbor nodes which have already received and replied to the Route Request packet (RREQ) and it's probable that they exist on other paths for the same source and destination, so even though they are located on two disjoint paths they will still affect each other in simultaneous data transfer. As mentioned in the abstract, our proposed algorithm tries to find zone-disjoint paths between source and destination. The nodes in zone-disjoint paths have almost no neighbor in the other path, to the feasible extent.

In brief, our proposed algorithm counts the number of active neighbors for each path from source to destination and eventually will choose paths that have the lowest total number of active neighbor nodes.

### A. *AODV Modifications*

In almost all the implementations of AODV algorithm, the intermediate nodes in a path will use a "Route Cache" table where they store the discovered paths. As a result, if a node receives a RREQ packet for which there exists a known path in the Route Cache table, the node will send a RREP message to the sender.

In ZD-AOMDV, there is no need for the intermediate nodes to have Route Cache tables and as a result the destination will receive all the path-request messages from different paths.

Also in the proposed algorithm, each node should save the RREQ messages it receives from other nodes in RREQ_Seen table, so that it can respond to the queries it receives from its neighbors. Also, there is an additional field in RREQ_Seen table of each node, called After_Active_Neighbor_Count (After_A_N_C), which will be used to count the number of active neighbors identified after sending the RREQ message.

In order to let the subsequent nodes to know the total number of active neighbors of the traversed nodes along a path, a new field called ActiveNeighborCount is added to the headers of both RREQ and RREP messages.

Also two new messages, RREQ_Query and RREQ_Query_Reply, are added to the route discovery process.

### B. *Proposed algorithm procedure*

Once a node intends to send data to a certain destination and doesn't find a valid path in its route table, it starts the route discovery process with broadcasting a RREQ packet to

[5] Ad-hoc On-Demand Multi-path Distance Vector.

[6] source neighbor nodes.

[7] Zone-Disjoint Ad-hoc On-demand Multi-path Distance Vector.

its neighbors and the ActiveNeighborCount field in the RREQ packet is set to zero.

Same as AOMDV, each neighbor upon receiving the RREQ packet will insert its name into the RREQ packet as the founder of one of probable paths and then will store the information of the packet in its reverse path table.

In the next step each path founder will send a RREQ_Query to its own neighbors asking them if they have already received the same RREQ packet before. Then each path founder will wait for neighbor's answer for a certain amount of time. Then each path founder will increase the ActiveNeighborCount field in the RREQ packet based on the number of positive responses it has received from its neighbors, and broadcast it.

At the same time, upon receiving the RREQ_Query message, each neighbor should search in the RREQ_Seen table and check if it has received the same RREQ message before or not and respond to this question with the RREQ_Query_Reply message. If this is the first time that it is receiving this RREQ, then the RREQ is also stored in the RREQ_Seen table along with the asking neighbor information.

Since in ZD-AOMDV the redundant RREQ messages are not thrown away, some nodes will receive the same RREQ message again and as a result they perform the query once more. So we should make sure that only the new neighbors will respond to this query. As a result, if a node receives a query message (RREQ_Query) for an existing RREQ in its RREQ_Seen table, it will respond with RREQ_Query_Reply message only if it has received the query from a new neighbor node, otherwise it will ignore the query.

Also it is possible that a node receives a query message from a new neighbor node corresponding to an existing RREQ which a query had already been performed for it before, and had already broadcasted that RREQ message. Since this new neighbor has not been considered in active node calculations, a new field called After_A_N_C is added to the RREQ_Seen table in each node. This field will be incremented once a node receives a query for an existing RREQ message from a new neighbor which had already been broadcasted. (Note that the receiving node will respond with the RREQ_Query_Reply in this case)

Each node will add the content of this field to the ActiveNeighborCount filed in the RREP message, once it receives the RREP message sent from destination to the source. This will ensure that the source upon receiving the RREP messages from destination can find out the exact number of active neighbors of each path.

Once the source node receives the first RREP packet, it will wait for a certain amount of time for other RREP packets to arrive from different paths. Then, it chooses the paths with least number of active neighbors for load-balanced transferring of data.

In AOMDV, as soon as the source has received the first RREP packet it begins transferring of data to the destination, which means that the path with least hops is used for data transfer. While in ZD-AOMDV, data transfer is postponed until several RREP packets are received. This helps the source node to choose the paths that are far from each other (zone disjoint paths).

### C. *Pseudo code of ZD-AOMDV*

The steps taken by source node, destination node and intermediate nodes are listed in figures 2, 3 and 4 respectively.

1. If there is data to be sent to a certain destination and there is no valid path for that destination, broadcast the RREQ packet.
2. Wait for first RREP packet to arrive.
3. After receiving the first RREP packet, wait for a certain amount of time, then from received paths choose those ones that have the least active neighbors and starts load balancing data transfer on these paths.

Figure 2. Pseudo code for the source node in ZD-AOMDV.

1. Send back a RREP packet to all the nodes from which a RREQ packet is received.

Figure 3. Pseudo code for the destination node in ZD-AOMDV.

1. Once a RREQ message is received and is acceptable (based on three aforesaid conditions in the Related Works section) perform the following steps:
   i. Save this message in the RREQ_Seen table
   ii. Construct the RREQ_Query packet
   iii. Send the RREQ_Query packet to the neighbors asking them if they have already seen the same RREQ message before
   iv. Wait for a certain amount of time for RREQ_Query_Reply messages from neighbors
   v. Increase the ActiveNeighborCount field based on the positive responses that are received
   vi. Broadcast the RREQ message with the new ActiveNeighborCount value.
2. Once a RREQ_Query message is received one of the following steps are performed:
   i. If based on the RREQ_Seen table this is a new RREQ message, store the RREQ in the RREQ_Seen table
   ii. If based on the RREQ_Seen table the same RREQ has been received from the same node, ignore the query.
   iii. If the RREQ is an existing one in the RREQ_Seen table and it has been received from a new neighbor but the same RREQ had not been broadcasted before, then only respond with sending back the positive RREQ_Query_Reply.
   iv. If the RREQ is an existing one in the RREQ_Seen table and it has been received from a new neighbor and the same RREQ had already been broadcasted, then respond with sending back the positive RREQ_Query_Reply and also increment the After_A_N_C field of the corresponding RREQ in the RREQ_Seen table.
3. Once the RREP message is received add the content of the corresponding After_A_N_C to the ActiveNeighborCount field of the RREP and send the RREP.

Figure 4. Pseudo code for the intermediate node in ZD-AOMDV

To better understand our proposed algorithm a hypothetical network is presented in figure 5-a as an example.

In figure 5-a, node S intends to transfer data to node D and it finds that there is no path in its routing table to the destination. It will initiate the route discovery process by broadcasting the RREQ packet to all its neighbors.

Nodes A, B, and C upon receiving the RREQ will insert their address into it as the founder of a path and will save the RREQ message in their RREQ_Seen table. Also they will reset the corresponding After_A_N_C value to zero in the RREQ_Seen table. Then each of them will broadcast RREQ_Query packet. After performing the queries, nodes A and C will each recognize node B as their active neighbor and will increment the ActiveNeighborCount in their RREQ message by one. Node B will recognize both A and C as active neighbors and will increment its ActiveNeighborCount by two. After that each of the nodes A, B and C will broadcast their RREQ packet.

The first RREQ message arrives to destination through node B. Up to this point the active neighbor value of this path (S-B-D) is two. Also nodes E and F will receive the RREQ message at his point and will initiate the query process, through which they recognize node B as their active neighbor and will increment the *ActiveNeighborCount* in their RREQ message. Also since node B has already broadcasted the same RREQ message before; it will respond to node's E and F query with positive *RREQ_Query_Reply* and at the same time will increment the *After_A_N_C* field of this RREQ in its *RREQ_Seen* table by one for each of those queries.

As shown in figure 5-b all the RREQ messages which arrive at destination will have the same ActiveNeighborCount value equal to two. In this figure the ActiveNeighborCount value (Left Number) of each RREQ at each node just before broadcasting the RREQ is shown along with the ActiveNeighborCount value (Right Number) of the RREQ at each node. As shown for node B, the value of After_A_N_C is equal to 2 for this RREQ.

The destination will respond back to its neighbors with a RREP message for each RREQ that it receives and will also update the ActiveNeighborCount value of RREP message with the same value in the RREQ. Each intermediate node on the path will add its After_A_N_C value to the ActiveNeighborCount value in the RREP message it receives. As shown in Figure 5-c, only node B will end up changing the ActiveNeighborCount value in the RREP and will increment it by two. Eventually the RREP messages will arrive at source.

After the source node receives the first RREP the source will allocate a certain time for other RREP messages to arrive. After the timer is expired the source will choose the paths which have lower ActiveNeighborCount values and will start sending data to destination through these zone-disjoint paths. In out example the source will choose the two paths shown in Figure 5-d which are S-A-E-D AND S-C-F-D for transferring data.

## IV. Performance Evaluation

In order to evaluate ZD-AOMDV, we have compared its performance to AOMDV with regards to several performance metrics.

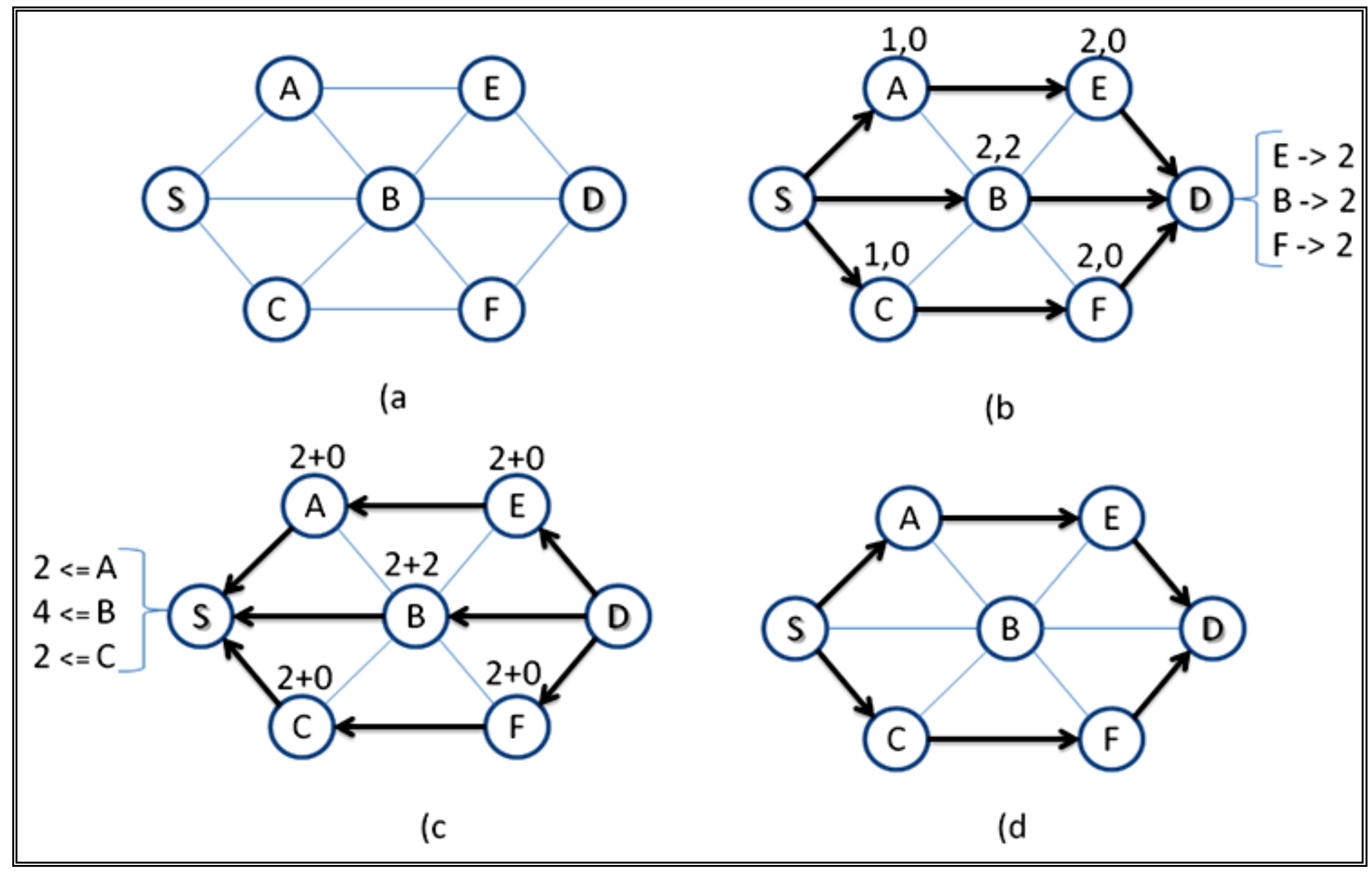

Figure 5. An example for proposed routing algorithm

*A. Simulation Environment*

We have used GloMoSim [11] as the simulation environment. In our scenario both algorithms use three paths for load-balanced transferring of data. Also, the area in which the nodes are spread is 1000×1000 meters and there are 100 nodes which can move in a range of 250 meters in random direction. The traffic model used for each node is CBR. Also in our simulation scenario each node uses the IEEE802.11 protocol in its MAC layer and for the purpose of sending or receiving data the standard RADIO-ACCNOISE model is used. In the random movement model chosen in this scenario, each node selects an arbitrary destination point and will remain still for a period of one second after it reaches destination. The total simulation time is 300 seconds and the results are the average of 25 times of simulation.

*B. Performance Metrics*

Three important performance metrics were evaluated in our simulation: (i) Average End-to-End Delay of packets – this includes all possible delays caused by buffering during route discovery phase, queuing at the interface queue, retransmission at the MAC layer, propagation and transfer delays – (ii) Packet Delivery Ratio, (iii) Control Overhead Ratio – the number of routing control packets in simulation time.

*C. Simulation Results*

*1) Packet Delivery Ratio*

By increasing each node's maximum speed, the packet delivery rate decreases in both algorithms, but the simulation result shown in figure 6 shows that ZD-AOMDV achieves higher rate of packet delivery compared to AOMDV. This is due to selection of zone-disjoint paths which decreases the collisions at the MAC layer.

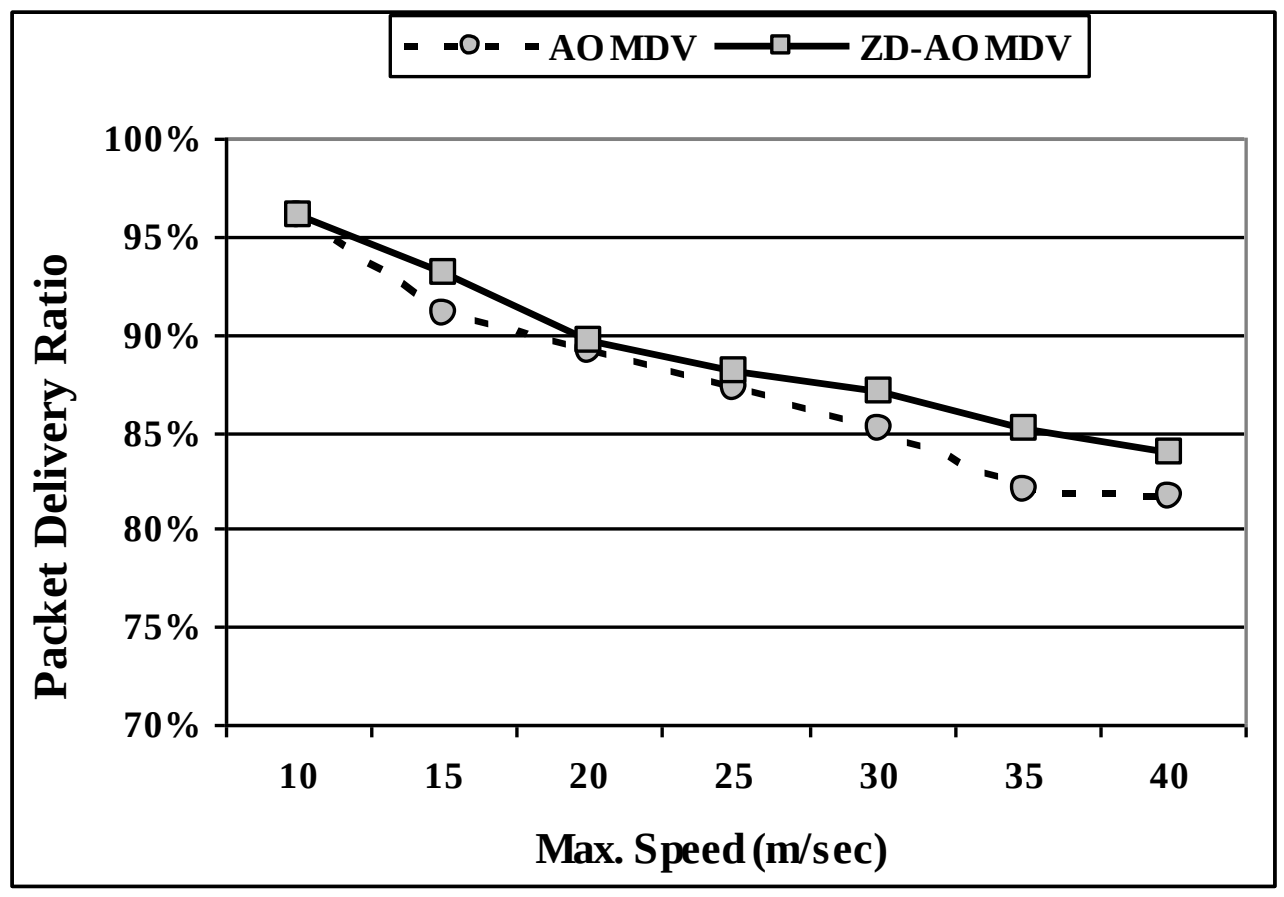

Figure 6. The packet delivery ratio with varying speed.

*2) Routing overhead*

In this section, the two algorithms are compared in terms of routing overhead. The maximum speed of each node is considered 25 meter per second. As shown in figure 7 the overhead of routing in ZD-AOMDV algorithm increases rapidly as the number of nodes increase. This is due to the increase in the number of query and query-reply packets sent between neighbors in the route discovery process.

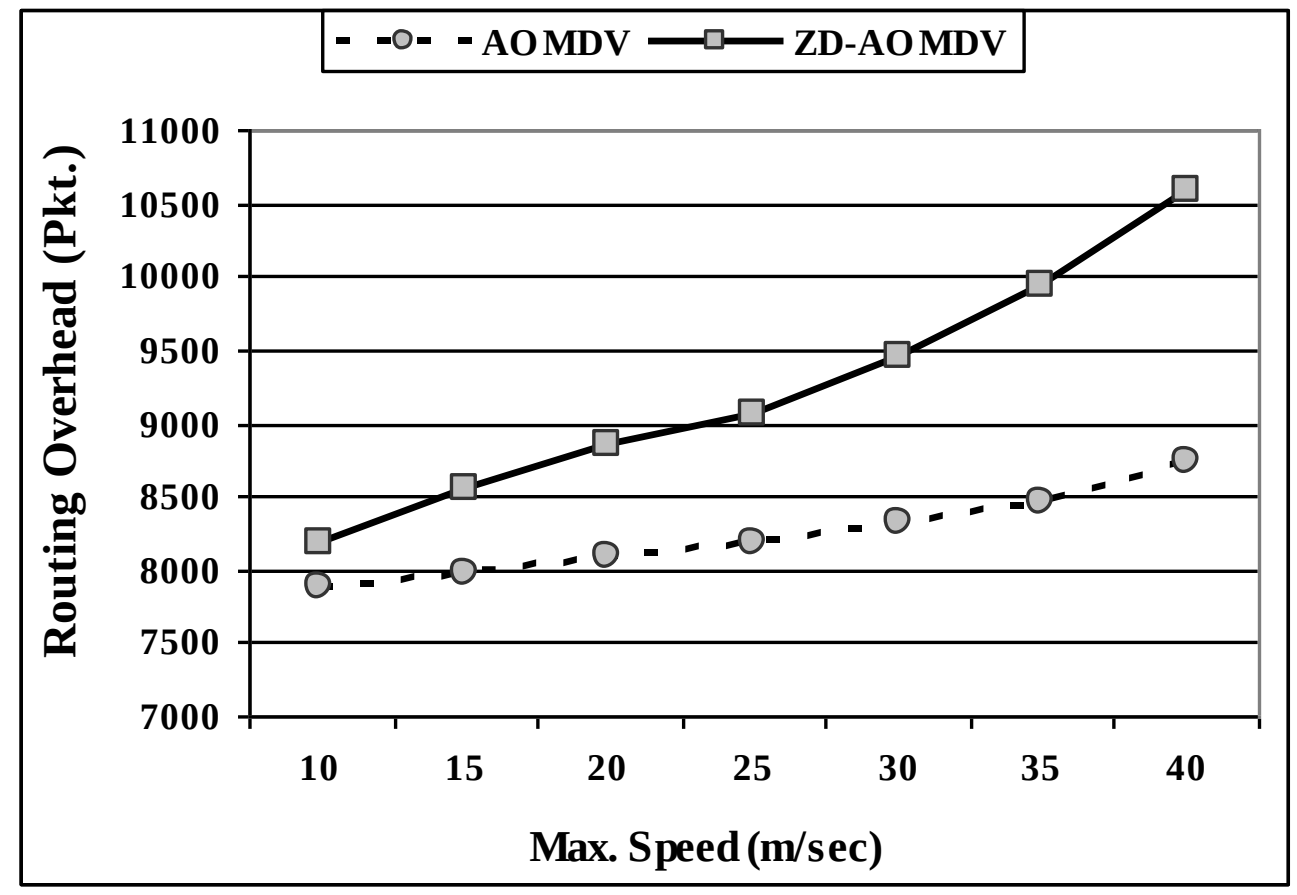

Figure 7. The routing overhead with varying speed.

*3) Average end to end delay*

Figure 8 illustrates simulation results for average end-to-end delay of our proposed algorithm based on nodes maximum speed compared to AOMDV. ZD-AOMDV achieves less average end to end packet delivery delay than AOMDV.

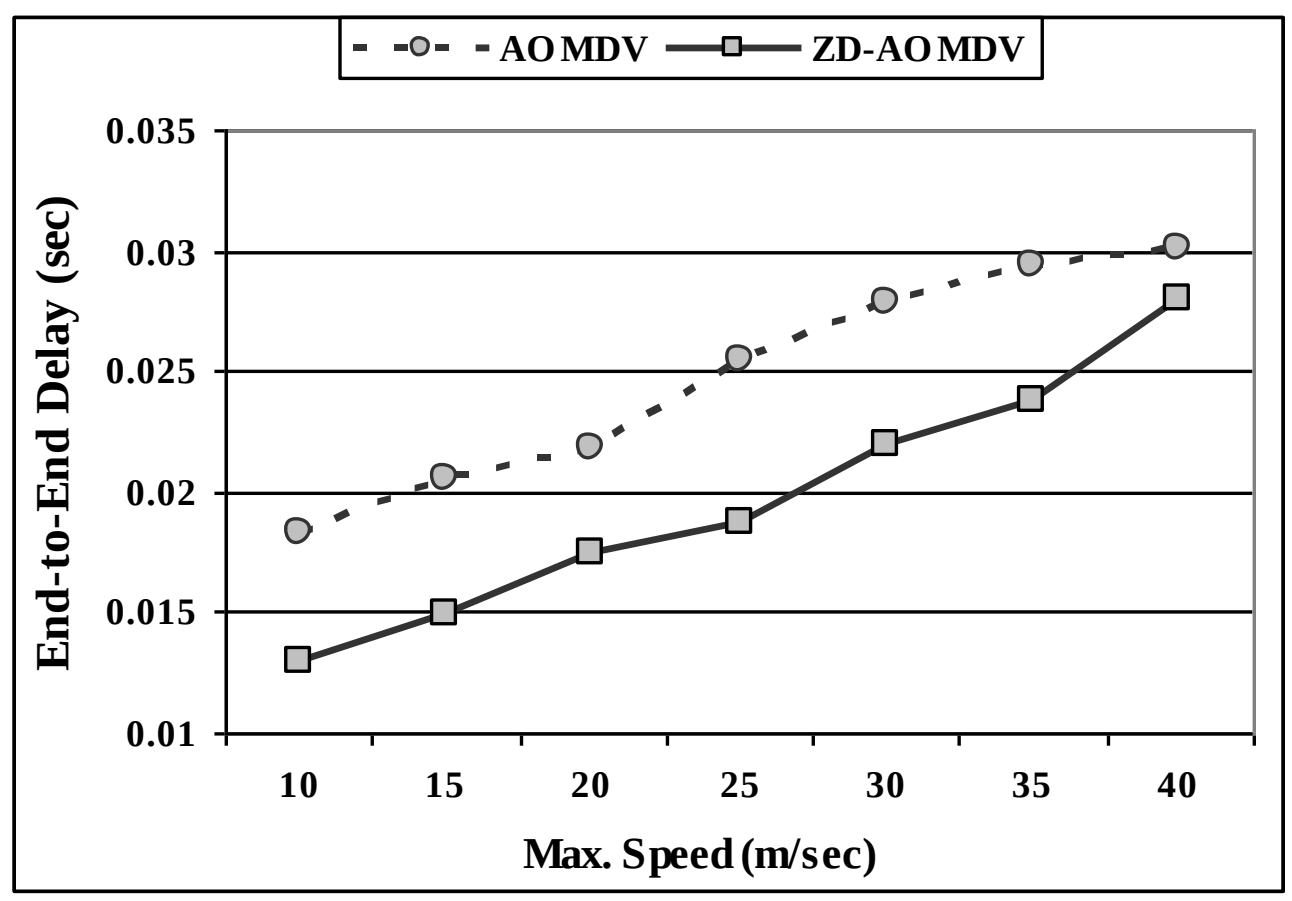

Figure 8. The average end-to-end delay with varying speed.

Although the Route Discovery phase of our proposed algorithm takes more time comparing to AOMDV but in data transfer phase ZD-AOMDV overtakes AOMDV and leads to less average end-to-end delay. This is due to selection of zone-disjoint paths.

## V. CONCLUSION

Multi-path routing algorithms in ad-hoc networks tend to use load balancing in order to transfer data between source and destination. This leads to less end-to-end delay since data is simultaneously transferred through several paths to destination. Although selecting node-disjoint paths for load balancing seems to be a good choice, but the nodes on these paths can still affect each other in the data transferring phase. This is because CSMA/CA protocol is used for acquiring channel access in wireless networks. To eliminate this problem we can utilize each node with directional antennas.

Figure 1. **Once a RREQ message is received and is acceptable (based on**

In this paper we proposed a multi-path protocol based on AODV which utilizes common omni-directional antennas, rather than directional ones and transfers data trough multiple zone disjoint paths simultaneously. For discovering zone-disjoint paths, the number of active neighbors in each path is calculated during the route discovery process. The total number of active neighbors for each path is the main parameter for selecting zone-disjoint paths. Finally, for evaluating our suggested protocol we compared it with AOMDV and achieved lower end-to-end delay and also increase in the packet delivery ratio.